\documentclass[reprint,prl,aps,twocolumn,superscriptaddress,longbibliography,preprintnumbers]{revtex4-2}
\usepackage[utf8x]{inputenc}
\usepackage{amssymb}
\usepackage{amsmath}
\usepackage{graphicx}
\usepackage{bbm}
\usepackage{psfrag}
\usepackage{latexsym}
\usepackage{color}
\usepackage[dvipsnames]{xcolor}
\usepackage{hyperref}
\hypersetup{colorlinks=true, citecolor=blue, urlcolor=blue, linkcolor=blue,urlcolor=cyan}
\usepackage{tensor}
\usepackage{amsfonts}
\usepackage{amsmath}
\allowdisplaybreaks[4]        
\usepackage{amssymb}
\usepackage{euscript}           
\usepackage{tensor}        
\usepackage{amsthm} 
\usepackage{graphicx}
\usepackage{subfigure}
\usepackage{bbm}
\usepackage[header,title,page,titletoc]{appendix}  
\usepackage[most]{tcolorbox}
\tcbuselibrary{skins, breakable, theorems}
\usepackage{verbatim}
\usepackage{cancel}

\newcommand{\ie}{{\it i.e.,}\ }

\newcommand{\mt}[1]{\textrm{\tiny #1}}

\renewcommand{\[}{\left[}

\newcommand{\sech}{\text{sech}}

\begin{document}
\preprint{
IFT-UAM/CSIC-25-123}

\author{Ben Craps}

\affiliation{\it Theoretische Natuurkunde, Vrije Universiteit Brussel (VUB) and The International Solvay Institutes, Pleinlaan 2, B-1050 Brussels, Belgium}

\author{Gabriele Pascuzzi}
\affiliation{\it Theoretische Natuurkunde, Vrije Universiteit Brussel (VUB) and The International Solvay Institutes, Pleinlaan 2, B-1050 Brussels, Belgium}

\author{Juan F. Pedraza}
\affiliation{\it Instituto de F\'isica Te\'orica UAM/CSIC, Calle Nicol\'as Cabrera 13-15, 28049 Madrid, Spain}
\author{Le-Chen Qu}
\affiliation{\it Instituto de F\'isica Te\'orica UAM/CSIC, Calle Nicol\'as Cabrera 13-15, 28049 Madrid, Spain}
\affiliation{\it Departamento de F\'isica Te\'orica, Universidad Aut{\'o}noma de Madrid, 28049 Madrid, Spain}
\author{Shan-Ming Ruan}
\affiliation{\it School of Physics, Peking University, Beijing 100871, China}
\affiliation{\it Center of High Energy Physics, Peking University, Beijing 100871, China}

\title{Explicit Connections Between Krylov and Nielsen Complexity}

\begin{abstract}
We establish a direct correspondence between Krylov and Nielsen complexity by choosing the Krylov basis to be part of the elementary gate set of Nielsen geometry and selecting a Nielsen complexity metric compatible with the Krylov metric. Up to normalization, the Krylov complexity of a Hermitian operator then equals the length squared of a straight-line trajectory on the manifold of unitaries that connects the identity operator with a precursor operator. The corresponding length provides an upper bound on Nielsen complexity that saturates whenever the straight line is a minimal geodesic. While for general systems we can only establish saturation in the limit of small precursors, we provide evidence that in broad classes of models and for suitable initial operators there is a precise correspondence between Krylov complexity and (the square of) Nielsen complexity for a finite range of precursors.
\end{abstract}

\maketitle

\noindent \textbf{Introduction.}
Quantum complexity has become a central concept at the intersection of quantum information, condensed-matter theory, and high-energy physics. One of the most influential approaches is \emph{Nielsen complexity} 
\cite{Nielsen:2005mkt,Nielsen:2006cea,Dowling:2006tnk}, which reformulates circuit complexity as a geometric problem. In this framework, a circuit corresponds to a curve on the unitary group manifold, and the complexity is given by the length of the minimal geodesic connecting the chosen unitary operator with the identity operator \cite{Jefferson:2017sdb,Khan:2018rzm,Hackl:2018ptj,Chapman:2018hou,Bhattacharyya:2018bbv,Guo:2018kzl,Bernamonti:2019zyy,Bernamonti:2020bcf,Caceres:2019pgf,Ruan:2020vze,Haque:2024ldr}. A complementary measure, \emph{Krylov complexity} 
\cite{Parker:2018yvk}, quantifies operator growth under Heisenberg evolution. Starting from an initial operator, one constructs an orthonormal Krylov basis via the Lanczos algorithm, thereby mapping the dynamics to a one-dimensional hopping problem along the Krylov chain. Krylov complexity directly measures the spread of the initial operator in Krylov space and provides a quantitative probe of operator growth, scrambling, and chaos \cite{Barbon:2019wsy,Avdoshkin:2019trj,Rabinovici:2020ryf,Jian:2020qpp,Dymarsky:2021bjq,Hornedal:2022pkc,Balasubramanian:2022tpr,Erdmenger:2023wjg,Caputa:2024vrn,Baggioli:2024wbz,Craps:2024suj,Huh:2024ytz}. For comprehensive reviews, see~\cite{Nandy:2024htc,Baiguera:2025dkc,Rabinovici:2025otw}.

Despite their different origins, Nielsen complexity and Krylov complexity both have deep ties to holography. Nielsen's geometric framework has motivated holographic proposals linking circuit complexity to geometric quantities that capture the growth of the black hole interior \cite{Susskind:2014rva,Stanford:2014jda,Brown:2015bva,Brown:2015lvg,Cai:2016xho,Belin:2021bga,Belin:2022xmt,Carrasco:2023fcj,Jorstad:2023kmq,Jiang:2023jti,Caceres:2023ziv,Myers:2024vve,Arean:2024pzo,Jiang:2025qai,Miyaji:2025yvm,Miyaji:2025jxy,Caceres:2025myu}. Meanwhile, in models of two-dimensional gravity, Krylov state complexity provides a concrete realization of the proposed correspondence between complexity and size of black hole interiors \cite{Lin:2022rbf,Rabinovici:2023yex,Heller:2024ldz,Balasubramanian:2024lqk,Aguilar-Gutierrez:2024nau,Xu:2024gfm,Heller:2025ddj,Ambrosini:2025hvo,Fu:2025kkh}. Although the possible relation between Nielsen complexity and Krylov complexity has attracted significant recent interest \cite{Lv:2023jbv,Caputa:2021sib,Craps:2023ivc,Aguilar-Gutierrez:2023nyk}, a direct and general correspondence remains elusive. 
Ref.~\cite{Craps:2023ivc} showed that Nielsen complexity and Krylov state complexity can be expressed as distinct functions of a common $q$-matrix, thereby revealing a formal similarity between their mathematical structures. 
However, this relation is neither an equality nor an inequality.
In the present paper, we show that the Krylov complexity of an operator is explicitly related to the Nielsen complexity of an associated precursor operator via an inequality that saturates in broad classes of models.

\vspace{4 pt}
\noindent \textbf{Mapping Krylov to Nielsen complexity.}
Krylov complexity characterizes the dynamical growth of a \emph{Hermitian} operator under time evolution. We begin with a normalized initial operator $\mathcal{O}(t=0)\equiv\mathcal{O}_0$, defined with respect to a chosen inner product, such as the infinite-temperature Wightman inner product,
\begin{equation}\label{eq:innerpro}
(\mathcal{O} | \mathcal{O}') := \frac{1}{\operatorname{Tr}[1]}\operatorname{Tr}\left[ \mathcal{O}^\dagger \mathcal{O}'\right] \,. 
\end{equation}
The Heisenberg evolution of the operator under a \emph{time-independent} Hamiltonian $H$, $\mathcal{O}(t)=e^{iHt}\mathcal{O}(0)e^{-iHt}$, is naturally formulated in operator space using the Liouvillian superoperator $\mathcal{L}\equiv[H,\cdot]$. Successive applications of $\mathcal{L}$ to the initial operator generate the Krylov subspace $\mathcal{K}=\mathrm{span}\{\mathcal{O}_0,i\mathcal{L}\mathcal{O}_0,(i\mathcal{L})^2\mathcal{O}_0,\dots\}$. An orthonormal basis $\{|\mathcal{O}_n)\}$ for this subspace can be constructed recursively via the Lanczos algorithm, \ie by applying the Gram–Schmidt procedure to the Krylov sequence $\mathcal{K}$. Using this orthonormal and Hermitian basis, the time-evolved operator can be expanded as
\begin{equation}\label{eq:operexpand}
|\mathcal{O}(t)) \equiv e^{i\mathcal{L}  t} |\mathcal{O}(0))  = \sum_{n}^{}  \varphi_n(t) | \mathcal{O}_n) \,, 
\end{equation}
with the initial condition $\varphi_n(0)=\delta_{n0}$ \footnote{Unlike some conventions in the literature, we absorb the factor $i^n$ into the basis $|\mathcal{O}_n)$ so that each element is manifestly Hermitian.}. Krylov complexity is thus defined as \cite{Parker:2018yvk,Balasubramanian:2022tpr,Chen:2024imd}
\begin{equation}\label{eq:defineCK}
  C_{\mt{K}}(\mathcal{O}(t)) := \sum_{m,n} f_{mn}\, \varphi_m(t) \varphi_n(t)\,, 
\end{equation}
where the positive matrix $f_{nm}$, referred to as the \emph{Krylov metric}, sets the weights for operator growth. A standard choice is $f_{mn}=f(n)\,\delta_{mn}$ with $f(n)=n$ \footnote{Since $f(0)=0$, the matrix is only positive semidefinite; we set $f(0)=\delta$ and take $\delta\to0$ at the end. More generally, any monotonically increasing $f(n)$ is admissible \cite{Balasubramanian:2022tpr}.}, 
yielding a natural measure of spreading along the Krylov chain. Another interesting choice, consistent with the conjecture that the growth rate of Nielsen complexity should be proportional to operator size \cite{Susskind:2014jwa,Roberts:2014isa}, is the Krylov metric $f_{mn}=n^{2}\delta_{mn}$, as discussed in part~A of the Supplemental Material.


In contrast, Nielsen complexity is defined for \emph{unitary} transformations, which mimic sequences of elementary unitary gates. 
The central idea of Nielsen's geometric formalism is to interpret the construction of a unitary operator $U(s)$ as a smooth trajectory on the manifold of unitaries, with control Hamiltonian (Hermitian velocity)
\begin{equation}
\label{trajectoryunitar}
H_{\mathrm{c}}(s)=i \partial_sU U^{-1}\,, \,\,  H_{\mathrm{c}}(s )\equiv Y^I(s) T_I \,. 
\end{equation}
This Hamiltonian is expanded in a basis of Hermitian generators $\{T_I\}$ representing the elementary gates, with control functions $Y^I(s)$ playing the role of tangent vectors along the trajectory. Taking the initial point as the identity, $U(0)=\mathbbm{1}$, and the endpoint as the target unitary, $U(1)=U_{\mt T}$, Nielsen complexity is then defined as the minimal length (also known as minimal ``cost'') among all admissible paths,
\begin{equation}\label{eq:defineCN}
C_{\mt{N}} (U_{\mt{T}}) = \min_{Y^I(s)} \int_0^1 \,\sqrt{\sum_{I,J}  G_{IJ}Y^I(s) Y^J(s)}\,ds \,.
\end{equation}
Here, the positive, right-invariant ``complexity metric'' (or ``cost function'') $G_{IJ}$ encodes the weights assigned to each gate, also known as ``penalty factors''. As a result, minimizing the cost is equivalent to solving the geodesic equation (or Euler–Arnold equation). 


A natural bridge between these two notions of complexity can be established by focusing on a special class of unitaries known as \emph{precursors} \cite{Susskind:2013lpa}, defined as
\begin{equation}\label{eq:precur}
 U_{\mt{T}}= e^{i H t} e^{-i z \mathcal{O}(0)} e^{-i H t} =e^{-i z \mathcal{O}(t)} \,,
\end{equation}
where $\mathcal O(t)$ is normalized using \eqref{eq:innerpro} and the positive parameter $z$ is the norm of the Hermitian generator $z\,\mathcal O(t)$ of $U_{\mt T}$. Equation~\eqref{eq:precur} links the time-evolved Hermitian operator $\mathcal O(t)$ that enters Krylov complexity to $U_{\mt T}$ in Nielsen's formalism. 
As we show below, the parameter $z$ will play a central role in relating the two notions of complexity. It is now straightforward to connect the identity to the target unitary via the straight-line trajectory  $U(s)=e^{-is H_{\mathrm{con}}}$, which corresponds to a constant control Hamiltonian,
\begin{equation}\label{eq:Hcon}
H_{\mathrm{c}}(s)  \rightarrow H_{\mathrm{con}} \equiv z \mathcal{O}(t) = z \sum_{n}^{}  \varphi_n(t) \mathcal{O}_n \,.
\end{equation}
The next step is to choose the basis of Hermitian generators $\{T_I\}$ used in Nielsen's construction to contain the Krylov basis $\{|\mathcal{O}_n)\}$, \ie $\{\mathcal O_n\}\subseteq\{T_I\}$. A simple correspondence is then realized by identifying
\begin{equation}\label{eq:gatechoose}
T_n \sim  \mathcal{O}_n \,, \quad Y^n \sim z \varphi_n \,.  
\end{equation}
Because the path $U(s)$ is a straight line with constant controls, its cost can be easily read off once the complexity metric is specified:
\begin{equation}
\operatorname{Cost}(\mathcal{O}(t)) = z\sqrt{\sum_{m,n} G_{mn}\,\varphi_m\varphi_n} \ge C_{\mt{N}} (U_{\mt{T}}) \,,
\end{equation}
which provides an upper bound on the Nielsen complexity of the precursor $U_{\mt T}$. Choosing $G_{IJ}$ such that $G_{mn}=f_{mn}$ for the directions corresponding to the Krylov basis, the two measures obey 
\begin{equation}\label{eq:inequality}
C_{\mt{N}} ( e^{-i z \mathcal{O}(t)} ) \le z \,\sqrt{C_{\mt{K}}(\mathcal{O}(t))} \,. 
\end{equation}
In fact, for targets close to the identity ($z\ll1$) \cite{Ali:2019zcj,Qu:2021ius,Qu:2022zwq}, the leading-order expansion yields \footnote{Higher-order corrections in this expansion were previously discussed in Ref.~\cite{Haque:2024ldr} and can be directly applied to the precursor considered here, as explained in part~C of the Supplemental Material.}
\begin{equation}\label{eq:premainresult}
    C_{\mt{N}}(e^{-i z \mathcal{O}(t)}) 
= z \sqrt{C_{\mt{K}}(\mathcal{O}(t))} + O(z^3)\,,
\end{equation}
explicitly linking the two notions of complexity for ``small'' precursor operators in general systems. Geometrically, this reflects the fact that any Riemannian manifold is locally flat, so the minimal geodesic connecting two sufficiently nearby points is well-approximated by the straight-line trajectory. 
Since the Krylov basis 
minimizes the cost function~\eqref{eq:defineCK} among all bases with an increasing sequence of penalty factors~\cite{Balasubramanian:2022tpr,Craps:2024suj}, it follows immediately from Eq.~\eqref{eq:premainresult} that the Krylov basis also minimizes the Nielsen complexity of ``small'' precursors. This makes the inclusion \eqref{eq:gatechoose} of the Krylov basis in the elementary gate set natural.

The remainder of this work addresses the question of whether the saturation of \eqref{eq:inequality} extends to finite $z$, \ie beyond the infinitesimal precursor limit. In other words, is the straight-line trajectory the shortest path from the identity operator to the target unitary $U_{\mt T}$? While for generic systems a trajectory generated by a constant control Hamiltonian $H_{\rm con}$ is not a geodesic, we will find that, for appropriate initial operators, the geodesic equation is satisfied for a large class of spin chains as well as for any physical fermionic system.
For the inequality~\eqref{eq:inequality} to be saturated,
\begin{equation}\label{eq:mainresult}
 C_{\mt{N}} ( e^{-i z \mathcal{O}(t)} ) = z \,\sqrt{C_{\mt{K}}(\mathcal{O}(t))}  \,, 
\end{equation}
it should moreover be checked that the geodesic is minimal.

To make the connection between the formalisms of Krylov and Nielsen complexities more explicit, we identify the inner product in both frameworks with Eq.~\eqref{eq:innerpro} and choose the gate basis $\{T_I\}$ so that $(T_I,T_J)=\delta_{IJ}$. The geodesic equation can then be recast in the compact form \cite{Dowling:2006tnk}
\begin{equation}\label{eq:geodesic}
\dot{H}_{\rm c} + i \, \mathbf{G}^{-1}\,([H_{\mathrm{c}}, \mathbf{G}(H_{\mathrm{c}})]) = 0 \,,
\end{equation}
where $\mathbf{G}(H_{\rm c}) =\sum_{I,J}G_{IJ} Y^J T_{I}$ \footnote{The chosen trace pairing $(T_I,T_J)=\delta_{IJ}$ naturally identifies the dual algebra $\mathfrak{g}^\ast$ with the Lie algebra $\mathfrak{g}$.}. It follows immediately that a constant control Hamiltonian, $\dot{H}_{\rm c}=0$, constitutes a geodesic whenever $\mathbf{G}^{-1}\!\left([H_{\mathrm{con}}, \mathbf{G}(H_{\mathrm{con}})]\right)=0$. We will now argue that this condition is satisfied for broad classes of models and suitable initial operators.

\vspace{4 pt}
\noindent \textbf{Models with straight-line geodesics.}
We begin with Majorana fermions obeying the Clifford algebra
\begin{equation}
\{\psi_i,\psi_j\}=\delta_{ij},\qquad i,j=1,\ldots,N,
\end{equation}
where $\{\cdot,\cdot\}$ denotes the anticommutator and $N$ is even. The space of linear operators acting on the $2^{N/2}$-dimensional Hilbert space, with operator basis $\{X_\alpha\}\equiv\{\psi_{i_1}\cdots\psi_{i_k}\}$ and $\alpha=1,\ldots,2^N$, naturally decomposes into two orthogonal sectors. To see this, introduce the chirality operator
\begin{equation}
    \psi^5 = (2i)^{N/2}\psi_1\psi_2\cdots\psi_N\,,
\end{equation}
which satisfies $(\psi^5)^2=\boldsymbol{1}$ and $\{\psi^5,\psi_i\}=0$. It then follows that the trace of a Majorana string vanishes for odd $k$, namely,
\begin{equation}
\operatorname{Tr}(\psi_{i_1}\cdots\psi_{i_k})=0\qquad (k\ {\rm odd}).
\end{equation}
Accordingly, the operator space can be decomposed into two sectors, $X_{\rm odd}$ and $X_{\rm even}$, with schematic algebraic structure
\begin{equation}\label{eq:lieal}
[X_{\rm odd}, X_{\rm odd}] \sim X_{\rm even},\qquad
[X_{\rm odd}, X_{\rm even}] \sim X_{\rm odd}.
\end{equation}
Next, consider a general physical fermionic Hamiltonian. Due to the parity superselection rule \cite{Zimboras:2013ewi}, such a Hamiltonian necessarily belongs to the even sector, as the appearance of any term with an odd number of fermions would allow for faster than light communication across spatially separate regions. Therefore, if we choose an initial operator $\mathcal{O}_0$ belonging to the odd sector, its time-evolution is fully restricted to this sector due to Eq.~\eqref{eq:lieal}.
Adopting a block-diagonal ansatz for the complexity metric in these sectors,
\begin{equation}
\mathbf{G} =
\begin{pmatrix}
\mathbf{G}_{\mathrm{odd}} & 0 \\
0 & \mathbf{G}_{\mathrm{even}}
\end{pmatrix},
\end{equation}
we now examine whether the straight-line trajectory $H_{\rm c} = H_{\rm con} \in X_{\rm odd}$ satisfies the geodesic equation \eqref{eq:geodesic}.
We observe that a simple solution arises if the penalty factor is homogeneous across Krylov sub-directions:
$\mathbf{G}(H_{\mathrm{con}}) = \mathbf{G}_{\rm odd}(H_{\mathrm{con}}) \propto H_{\mathrm{con}} \,.$
For more general penalty factors, however, the commutator $[ H_{\mathrm{con}}, \mathbf{G}(H_{\mathrm{con}}) ]$ does not vanish.
Nevertheless, owing to the algebraic fact that $[ H_{\mathrm{con}}, \mathbf{G}(H_{\mathrm{con}})] \in X_{\mathrm{even}}$, substituting the constant $H_{\rm con}$ into the geodesic equation yields:
\begin{equation}
\mathbf{G}^{-1}\,([H_{\mathrm{c}}, \mathbf{G}(H_{\mathrm{c}})]) \sim 
 (G_{\rm{even}})^{-1} \, G_{\rm{odd}} \,.
\end{equation}
Here, the subspace indices and the structure constants of the unitary group are suppressed for simplicity.
One then finds that the constant solution $H_{\rm c}=H_{\rm con}$ asymptotically satisfies the geodesic equation in the limit $(G_{\rm{even}})^{-1} \, G_{\rm{odd}}\to 0$ (equivalently, $G_{\rm even}\to\infty$ for fixed $G_{\rm odd}$).
This indicates that the straight-line trajectory within the Krylov subspace becomes a geodesic in the full operator space whenever the even sector is heavily penalized, \ie when its contribution is effectively suppressed by a large penalty factor. With these penalty factors and an odd initial operator, the straight line therefore satisfies the geodesic equation in any physical fermionic Hamiltonian. We note that while we need to start with an odd seed operator $\mathcal{O}_0\in X_{\rm odd}$ for our construction to work, physical measurements of odd operators are themselves forbidden by superselection. However, the Krylov complexity of $\mathcal{O}_0$ is completely controlled by its autocorrelation function. This, in turn, depends solely on the even parity combination $\mathcal{O}_0(t) \mathcal{O}_0$ and is therefore a well-defined physical quantity.

Another class of systems that display the same algebraic structure is given by a subset of spin-$1/2$ chains. First, note that any spin chain model is equivalent to some fermionic model via the Jordan-Wigner transform. However, the superselection rule we discussed does not apply here, since the condition of locality is not required for these auxiliary (non-physical) fermions. To arrive at the algebra \eqref{eq:lieal}, we instead start by considering a global spin-flip operator:
\begin{equation}
    \mathcal{P}_x = \prod_{j=1}^N S_j^x\,.
\end{equation}
We use the $x$-direction for concreteness, but what follows can also be done for the other directions. With the usual spin algebra, it can be shown that:
\begin{equation}
    \mathcal{P}_x S_j^x=S_j^x \mathcal{P}_x\,,\hspace{0.5cm}\mathcal{P}_x S_j^y=-S_j^y \mathcal{P}_x\,,\hspace{0.5cm}\mathcal{P}_x S_j^z=-S_j^z \mathcal{P}_x.
\end{equation}
With these relations, it is easy to see that given any Pauli string, the operator $\mathcal{P}_x$ counts the total number of spin-$y$ and spin-$z$ operators in the string, since it will commute with it when this number is even, and anticommute when it is odd. With this assignment, the operator space can once again be decomposed into an even and an odd sector, which follow the algebra of Eq.~\eqref{eq:lieal}. As before, we need our Hamiltonian to lie in the even sector and the initial operator to lie in the odd sector. Some examples of common spin chain Hamiltonians belonging to this class are the longitudinal or the transverse field Ising chain, as well as the Heisenberg chain.

Having shown that the straight line is a geodesic in these models, we now turn to the question of whether it realizes the shortest-length trajectory.

\begin{figure}[t]
\centering
\includegraphics[width=3in]{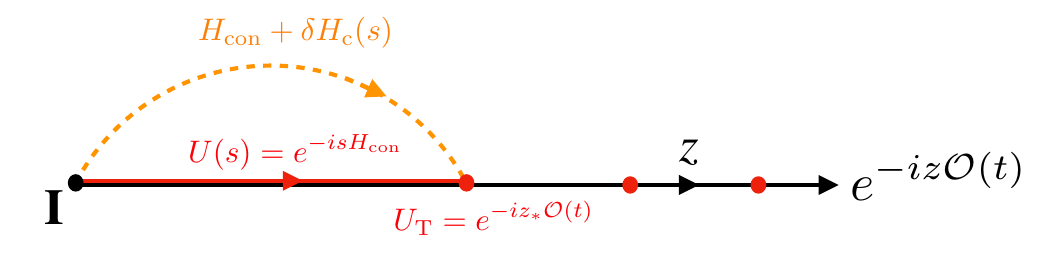}
\caption{Conjugate points (red dot) along the geodesic (black straight-line) are defined as the points where the perturbation satisfying the geodesic deviation equation vanishes.}
\label{fig:conjugatepoints}
\end{figure}

\vspace{4 pt}
\noindent \textbf{Minimal geodesics.} 
Nielsen complexity is defined as the minimal cost over all admissible paths, so it is essential to verify that the corresponding geodesic is indeed minimal, both locally and globally. A geodesic is locally length-minimizing if and only if it contains no conjugate points along its trajectory \cite{frankel2004geometry}, as illustrated in Fig.~\ref{fig:conjugatepoints}. 

To investigate the presence of such conjugate points in the models discussed above, we parameterize small perturbations around the reference trajectory as
\begin{equation}
\delta H_{\rm c}(s) = \delta H_{\rm o} (s) + \delta H_{\rm e} (s) \,. 
\end{equation}
Solving the perturbed geodesic equation with arbitrary penalty factors in these models is generally difficult. To gain intuition, we solve it for a special case, namely homogeneous penalties in the even and odd sectors. Furthermore, in part~B of the Supplemental Material, we solve the perturbed geodesic equation for the Berger sphere with arbitrary penalties. In all these cases, the location of the first conjugate point is bounded from below by~$\pi/2$, which can be taken to suggest that also in more general situations the first conjugate point will stay a finite distance away from $z=0$.

So, we assume homogeneous penalties in the even and odd sectors,
\begin{equation}\label{eq:homogeneous}
\mathbf{G}( X_{\rm even}) = G_{\rm even} \,  X_{\rm even}  \,,  \quad \mathbf{G}( X_{\rm odd}) = G_{\rm odd} \,  X_{\rm odd}    \,. 
\end{equation}
Under this simplification, the straight-line trajectory with a constant control Hamiltonian $H_{\rm con}=z\,\mathcal{O}(t)$ can be shown to satisfy the geodesic equation~\eqref{eq:geodesic}, thereby representing a geodesic in the space of unitaries. The associated geodesic deviation (Jacobi) equation governing small perturbations $\delta H_{\rm c}$ then takes the form \cite{Balasubramanian:2019wgd}
\begin{equation}
i \frac{d}{ds} \delta H_{\rm{o}} =  (\alpha-1) [H_{\rm{con}}, \delta H_{\rm{e}}], \qquad i \frac{d}{ds} \delta H_{\rm{e}} = 0 \,, 
\end{equation}
where we have introduced the ratio $\alpha \equiv G_{\rm even}/G_{\rm odd}$. After fixing the initial conditions, the corresponding solution of the geodesic deviation equation (\ie the Jacobi fields) can be expressed as
\begin{equation}\label{eq:JacobiSolution}
\delta H_{\rm c}(s) = \delta H_{\rm c}(0) - i (\alpha-1) \, s \, [H_{\rm{con}}, \delta H_{\rm{e}}(0)]\,. 
\end{equation}

Solutions for conjugate points in a closely related setup were analyzed in detail in Ref.~\cite{Balasubramanian:2019wgd}. 
The essential observation is that the first-order perturbation at the endpoint of the straight-line trajectory is given by
\begin{equation}\label{eq:pertubations1}
   (U^{-1} \delta U)|_{s=1} = \int_{0}^{1} d\tilde{s} \, e^{i  H_{\rm con} \tilde{s}} \delta H_{\rm c}(\tilde{s}) e^{-i H_{\rm con} \tilde{s}}  \equiv  \mathbf{\Delta} [ \delta H_{\rm c}]\,,
\end{equation}
which can be interpreted as the action of a superoperator $\boldsymbol{\Delta}$ on the perturbation $\delta H_{\rm c}$. 
The conjugate points are therefore determined by the zeros of this operator, \ie by the condition $(U^{-1} \delta U)|_{s=1}=0$. To evaluate the integral in Eq.~\eqref{eq:pertubations1} explicitly, we expand in the eigenbasis of the Hermitian operator $\mathcal{O}(t)$, defined by $\mathcal{O}(t)\,|n\rangle=\omega_n |n\rangle$ \footnote{The time dependence is encoded in the eigenstates as $|n\rangle\equiv|n(t)\rangle=e^{iHt}|n_0\rangle$, where $\mathcal{O}(0)\,|n_0\rangle=\omega_n|n_0\rangle$.}.
Assuming the initial perturbations take the form
\begin{equation*}
\delta H_{\rm c}(0) = \sum_{m,n} S_{mn} |m\rangle \langle n|\,,\quad \delta H_{\rm{e}}(0) = \sum_{m,n} M_{mn} |m\rangle\langle n| \,,
\end{equation*}
the integration in Eq.~\eqref{eq:pertubations1} yields
\begin{equation*}
\begin{split}
     \frac{(e^{i z \omega_{mn}} -1 ) S_{mn}  
      +  (\alpha-1) (e^{i z \omega_{mn}} (1 - i z \omega_{mn}) - 1) M_{mn} }{i\, z\, \omega_{mn}} \,,
     \end{split}
\end{equation*}
where $\omega_{mn} = | \omega_m - \omega_n |$. The solution for the geodesic
perturbation $\delta H_{\rm c}(s)$ in Eq.~\eqref{eq:JacobiSolution} is determined by an arbitrary initial perturbation spanning the operator space, \ie $\delta H_{\rm c}(0)=\sum_\alpha c_\alpha X_\alpha$. 
The existence of a conjugate point, $\boldsymbol{\Delta}[\delta H_{\rm c}]=0$, corresponds to the vanishing of the determinant:
\begin{equation}\label{eq:detDelta}
0= \det \mathbf{\Delta}_{\alpha \beta} \equiv  \det \left( \text{Tr} \left( X_\alpha \mathbf{\Delta}[X_\beta]  \right) \right) \,. 
\end{equation}
\begin{figure}[t]
\centering
\includegraphics[width=3.25in]{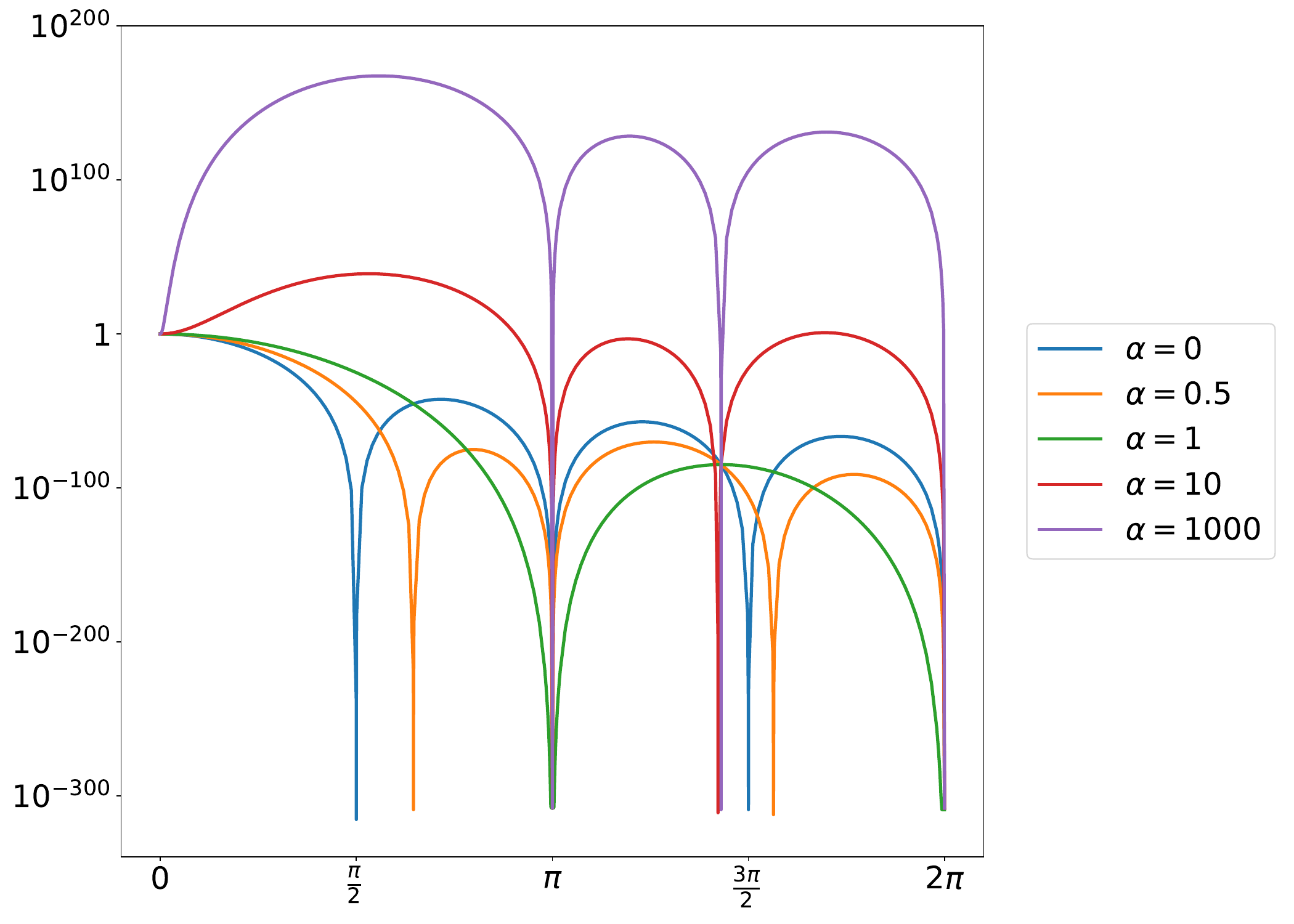}
\caption{Absolute value of the determinant in Eq.~\eqref{eq:detDelta} for the SYK model with $N = 8$, $q=4$, $\mathcal{J}=1$ and initial operator $\mathcal{O}(0)=\sqrt{2}\,\psi_1$, as a function of the precursor size $z$ for various penalty ratios $\alpha \equiv G_{\rm even}/G_{\rm odd}$. The same result would be found for other even Hamiltonians and odd initial operators, as long as the latter has eigenvalues $\pm1$.
}
\label{fig:DetY}
\end{figure}

For concreteness, we perform the numerics that follow in the SYK model \cite{Sachdev_1993,KitaevTalks}. (We obtained similar results for spin chains.) The model consists of $N$ Majorana fermions with all-to-all random interactions \cite{Maldacena:2016hyu}. Its Hamiltonian is
\begin{equation}
H = i^{q/2} \sum_{1 \leq i_1 < \cdots < i_q \leq N}
J_{i_1 \cdots i_q}\,\psi_{i_1}\cdots\psi_{i_q}\,,
\end{equation}
with $q$ taken to be even 
and where the couplings $J_{i_1\cdots i_q}$ are drawn from a Gaussian ensemble with vanishing mean and variance
\begin{equation}
\langle J_{i_1 i_2 \ldots i_q}^2\rangle
=
\frac{2^{q-1}\mathcal{J}^2 (q-1)!}{q\,N^{q-1}}\,.
\end{equation}
Here $\mathcal{J}$ sets the characteristic energy scale of the model. In what follows we set the parameters to $N=8$, $q=4$ and $\mathcal{J}=1$, and start with the initial operator $\mathcal{O}(0)=\sqrt{2}\,\psi_1$. Numerical evaluation of the determinant in Eq.~\eqref{eq:detDelta} for this setup is shown in Fig.~\ref{fig:DetY}. Although obtaining a closed-form expression for the exact locations of the conjugate points is challenging, it is straightforward to observe that, in the special case $\alpha=1$, corresponding to the bi-invariant complexity metric, the determinant vanishes at $z=\tfrac{2\pi k}{\omega_{mn}}$ for integer $k$. The first conjugate point therefore occurs at $z_\ast=\tfrac{2\pi}{\omega_{\max}}$, where $\omega_{\max}$ denotes the maximum spectral gap. Choosing $\mathcal{O}(t)=\sqrt{2}\,\psi_1(t)$ gives $\omega_{\max}=2$, and thus $z_\ast=\pi$. Intuitively, reducing the penalty assigned to even-sector operators (\ie decreasing $\alpha$) brings the conjugate points closer, whereas increasing $\alpha$ pushes them further away. This trend is confirmed by our numerical results. More importantly, the location of the first conjugate point is bounded below:
\begin{equation}
(\text{first conjugate point:}) \quad z_\ast(\alpha) \ge \frac{\pi}{2} \,,
\end{equation}
with the lower bound saturated at $\alpha=0$. In part~B of the Supplemental Material, we analytically verify that this bound also holds for the Berger sphere with arbitrary penalties. We thus find that, in all models that we investigated, the straight-line trajectory \eqref{eq:Hcon} remains a locally minimal geodesic for any precursor operator with $z \le \tfrac{\pi}{2} 
$. This motivates us to expect that, for the models discussed previously, with penalty schedules of interest, the straight-line trajectory \eqref{eq:Hcon} will be a locally minimal geodesic for a finite range of $z$.

Our conjugate-point analysis was simplified by the specific choice $\mathcal{O}_0=\sqrt{2}\psi_1$, for which $\mathcal{O}(t)^2=1$, so the spectrum of $\mathcal{O}(t)$ is fixed to $\pm 1$. 
For a more general odd initial operator, the conjugate-point locations, and hence the precise range of finite-$z$ saturation, may depend on the spectrum of $\mathcal{O}(t)$. We do expect a finite saturation interval in many such cases, even though its endpoint need not coincide with the one found for $\mathcal{O}_0=\sqrt{2}\psi_1$. In contrast, for even initial operators, we generally do not expect our inequality between Nielsen and Krylov complexity to saturate for finite $z$.

To complete our construction, we now discuss the global minimality of geodesics. Since the unitary group $\mathrm{SU}(N)$ is compact, geodesic loops can occur: beyond a certain distance, two distinct geodesics may connect the same endpoints with equal length. In the SYK model, starting from the Hermitian operator $\mathcal{O}(0)=\sqrt{2}\,\psi_1$, whose eigenvalues are fixed at $\pm1$, it is
straightforward to see that the geodesic forms a closed loop at $z=2\pi$, since
\begin{equation}
U_{\mt{T}} |_{z=2\pi}=\mathbbm{1} =  U(s=0)  \,. 
\end{equation}
Combining this with the local minimality condition, we conclude that the locally minimizing geodesic remains globally minimizing up to the midpoint $z=\pi$. In this regime, we establish an explicit correspondence between the Krylov complexity $C_{\mt K}(\mathcal{O}(t))$ and the Nielsen complexity $C_{\mt N}(e^{-iz\,\mathcal{O}(t)})$ by showing that the straight-line trajectory $U(s)=e^{-isz\,\mathcal O}$ is the globally minimizing geodesic underlying both measures of complexity.

We comment on the relation of our results with Ref.~\cite{Aguilar-Gutierrez:2023nyk}, which demonstrated that Krylov complexity does not satisfy the triangle inequality and therefore cannot be interpreted as a distance measure on the space of operators.
 First, we have shown that Krylov complexity is naturally interpreted as the square of a distance, rather than a distance itself -- a possibility that was also considered in Ref.~\cite{Aguilar-Gutierrez:2023nyk}. Second, in our prescription the distance does not involve the target operator itself, but rather a precursor associated with this operator. Third, and crucially, 
 $z \sqrt{C_{\mt{K}}(\mathcal{O}(t))}$ is mapped to a distance between the identity and the precursor of $\mathcal{O}(t)$. Since all distances in the Nielsen geometry are measured relative to the identity, our correspondence does not give rise to an obvious triangle construction that would require Krylov complexity to satisfy the triangle inequality.

\vspace{4pt}
\noindent \textbf{Summary.}
In this work, we have established a direct correspondence between Krylov complexity and Nielsen complexity by embedding the Krylov basis and weight metric into the elementary gate set and complexity metric of Nielsen's geometric approach. We showed that Krylov complexity is mapped to the square of the length of a straight-line trajectory on the manifold of unitaries, which generally serves as an upper bound on Nielsen complexity. The bound generally saturates in the limit of ``small'' precursors ($z\to 0$). We presented evidence that, for suitable initial operators, it also saturates for a finite range of $z$ in any physical fermionic model, as well as in a large class of spin chain models. Our results thus establish a concrete geometric bridge between two distinct yet complementary measures of quantum complexity.

\vspace{8pt}

\begin{acknowledgments}

\noindent \emph{Acknowledgments.}
We are grateful to Jos\'e Barb\'on, Hong-Yue Jiang, and Yu-Xiao Liu for valuable discussions and comments on the manuscript. Collaboration with Marine De Clerck, Oleg Evnin and Philip Hacker on related topics is also gratefully acknowledged. Work at VUB is supported by FWO-Vlaanderen projects G012222N and G0A2226N and by the VUB Research Council through the Strategic Research Program High-Energy Physics. GP is further supported by a PhD fellowship from FWO. JFP is supported by the Spanish MINECO ‘Centro de Excelencia Severo Ochoa' program under grant SEV-2012-0249, the Comunidad de Madrid `Atracci\'on de Talento' program (ATCAM) grant 2020-T1/TIC-20495, the Spanish Research Agency via grants CEX2020-001007-S and PID2021-123017NB-I00, funded by MCIN/AEI/10.13039/501100011033, and ERDF `A way of making Europe.' LCQ is funded by a scholarship from the China Scholarship Council (CSC) and acknowledges support from a short-term scientific mission grant under COST Action CA22113 THEORY-CHALLENGES. LCQ also thanks the Vrije Universiteit Brussel (VUB) for hospitality while part of this work was carried out. SMR is supported by Peking University under startup Grant No. 7101303985, and gratefully acknowledges the hospitality of the Instituto de Física Teórica (IFT UAM/CSIC) and Beijing Normal University while part of this work was carried out. 
\end{acknowledgments}

\bibliography{References}

\pagebreak

\appendix

\setcounter{equation}{0} 
\renewcommand{\theequation}{S\arabic{equation}} 

\titlepage

\setcounter{page}{1}

\begin{center}
{\LARGE Supplemental Material}
\end{center}

\section{A. SYK model and operator size}\label{sec:appB}
In the main text, we demonstrated that the straight-line control $H_{\rm con} = z\mathcal{O}(t)$ satisfies the geodesic equation~\eqref{eq:geodesic} in the SYK model. Of particular interest is the large-$N$ regime with $N \gg q^2 \gg 1$.
In this limit, the $n$-th Krylov basis element $\mathcal{O}_n$ can be expressed as a superposition of Majorana strings of fixed length~\cite{Roberts:2018mnp,Parker:2018yvk,Bhattacharjee:2022lzy},
\begin{equation}
\mathcal{O}_n = i^n \sum_{i_1 < \cdots < i_s} c_{i_1 \cdots i_s}\, \psi_{i_1} \cdots \psi_{i_s} + O\left(\frac{1}{q}\right)\,,
\end{equation}
where the operator size is $s = n(q-2) + 1$, with $n$ labeling the Krylov generation~\cite{Roberts:2018mnp,Parker:2018yvk}.  
Correspondingly, the associated Krylov wavefunctions admit compact analytic forms~\cite{Roberts:2018mnp,Parker:2018yvk}:
\begin{equation}
\varphi_n(t) =
\begin{cases}
1 + \dfrac{2}{q}\, \ln \sech(\mathcal{J} t) + O\left(\tfrac{1}{q^2}\right), & n=0, \\[6pt]
\sqrt{\dfrac{2}{nq}}\, \tanh^n(\mathcal{J} t) + O\left(\tfrac{1}{q^2}\right), & n \geq 1\,.
\end{cases}
\end{equation}
The average operator size then follows as
\begin{equation}
s(t) = \sum_n \left[1 + (q - 2)n \right] \varphi_n^2(t)
      = \cosh(2\mathcal{J} t) + O\!\left(\tfrac{1}{q^2}\right)\,.
\end{equation}

A particularly illuminating choice of the Krylov metric arises when adopting a quadratic form,
\begin{equation}
f_{nm}= n^2 \, \delta_{nm} \,,
\end{equation}
in contrast to the more common linear choice $f_{nm} = n\,\delta_{nm}$.
This quadratic weighting leads to a Nielsen complexity of the form
\begin{equation}
C_{\mt{N}} ( e^{-i z \mathcal{O}(t)} ) = z\sqrt{\sum_n n^2 \varphi_n^2(t)}
           = z\sqrt{\tfrac{1}{2q}}\, \sinh(2\mathcal{J} t)
           +  O\!\left(\tfrac{1}{q^2}\right) \,. 
\end{equation}
Taking the time derivative thus yields 
\begin{equation}\label{ckdot}
\frac{dC_{\mt{N}} ( e^{-i z \mathcal{O}(t)} )}{dt}
  = z \mathcal{J} \sqrt{\tfrac{2}{q}}\, s(t)
  + \mathcal{O}\!\left(\tfrac{1}{q^2}\right) \,. 
\end{equation}
This result naturally realizes the conjecture proposed in \cite{Susskind:2014jwa,Roberts:2014isa} that the growth rate of quantum complexity is proportional to the operator size.


\section{B. Jacobi fields on Berger sphere}\label{sec:appD}

In the main text, we analyzed the appearance of conjugate points within the finite-$N$ SYK model under the assumption of homogeneous penalty factors between the even and odd sectors, \ie eq.~\eqref{eq:homogeneous}.
This simple choice of penalty factors facilitates the solution of the Jacobi equations on a high-dimensional group manifold. However, it also leads to a trivial evolution of Krylov complexity, since all Krylov basis elements are assigned identical weights. 

As solving for the conjugate points with fully arbitrary penalty factors remains analytically challenging, here we consider another simplified model that captures essential features of the problem. Specifically, we focus on a single-qubit system equipped with inhomogeneous penalty factors. Because this system is parameterized by a three-dimensional manifold, the Jacobi equations can be solved exactly, allowing us to determine the location of the conjugate points explicitly. We find that the result agrees with the simplified analysis presented in the main text: the location of the first conjugate point remains bounded from below by $\pi/2$.

For a single qubit, the Nielsen geometry of complexity corresponds to the group manifold $SU(2)$ with a right-invariant metric, often referred to as the Berger sphere \cite{Brown:2019whu}. Despite its simplicity, this setting captures essential features of geodesic deviation and offers exact solutions to the Jacobi equation. We can parametrize $SU(2)$ using Euler angles $\theta_{x,y,z}$ as
\begin{equation}
U = e^{i \sigma_z \theta_z} e^{i \sigma_y \theta_y} e^{i \sigma_x \theta_x},
\end{equation}
where the Pauli matrices $\sigma_{x,y,z}$ generate the Clifford algebra $\mathbb{CL}_2$ via
\begin{equation}
\sigma_x = \sqrt{2} \psi_1 \,, \sigma_y = \sqrt{2} \psi_2 \,, \sigma_z = -2i \psi_1 \psi_2 \,.
\end{equation}
This is the simplest algebra realizing the particular algebra \eqref{eq:lieal} by identifying 
\begin{equation}
X_{\rm odd} = \{  \sigma_x , \sigma_y\} \,, \qquad  X_{\rm even}= \{ \sigma_z  \}\,. 
\end{equation}

Let us first consider the most general right-invariant metric on this manifold by assigning independent penalty factors $P_x$, $P_y$, and $P_z$ to each direction. Fixing $P_x = 1$ without loss of generality, the metric takes the form~\cite{Brown:2019whu}
\begin{equation}
g_{i j}=\left(\begin{array}{ccc}
\cos^2 (2\theta_y)(\cos^2 (2\theta_z) + P_y \sin^2 (2\theta_z)) + P_z \sin^2 (2\theta_y) & \frac{1}{2}\cos(2\theta_y) \sin (4\theta_z) (1-P_y) & 2P_z \cos \theta_y \sin \theta_y \\
\frac{1}{2}\cos (2\theta_y) \sin (4\theta_z) (1-P_y) & \frac{1}{2} (P_y + 1 + (P_y - 1)\cos (4\theta_z) ) & 0 \\
2P_z \cos \theta_y \sin \theta_y & 0 & P_z
\end{array}\right).
\end{equation}
For the simplest case with $P_x=P_y=P_z$, this gives rise to
\begin{equation}
 \left.d s^2\right|_{P_x=P_y=P_z}=\cos ^2 (2 \theta_y) d \theta_x^2+d \theta_y^2+  \left(d \theta_z+\sin (2 \theta_y) d \theta_x\right)^2 \,. 
\end{equation}
Using the coordinate transformation
\begin{equation}
\phi=  \theta_x- \theta_z \,, \quad \cos \psi = \cos \theta_y \cos(\theta_x + \theta_y) \,, \quad \tan \theta= \frac{\sin \theta_y}{\cos \theta_y \sin (\theta_x + \theta_z)} \,,
\end{equation} 
we can find that the corresponding complexity geometry is nothing but the standard round metric on a three-sphere:
\begin{equation}
 \left.d s^2\right|_{P_x=P_y=P_z}= d \psi^2+\sin ^2 \psi\left(d \theta^2+\sin ^2 \theta d \phi^2\right) \,.
\end{equation}

\begin{figure}[t]
    \centering
    \includegraphics[width=4in]{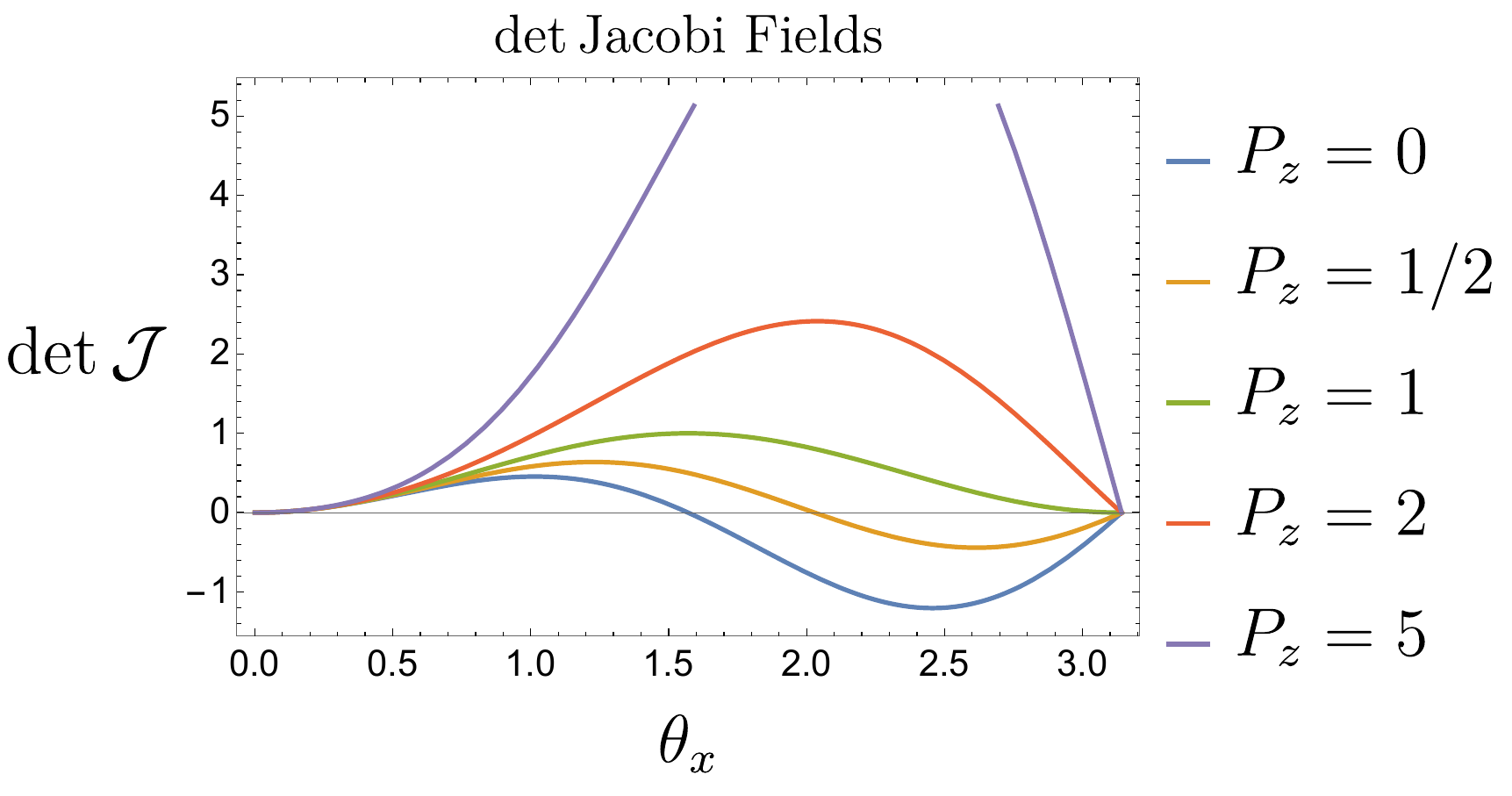}
    \caption{Jacobi fields and conjugate points on Berger sphere with odd generators $P_x=P_y=1$ and different values of the even generator $P_z$. Conjugate points appear when $\det \mathcal{J}=0$.}\label{fig:DetJ}
\end{figure}

\begin{figure}[t]
    \centering
    \includegraphics[width=4in]{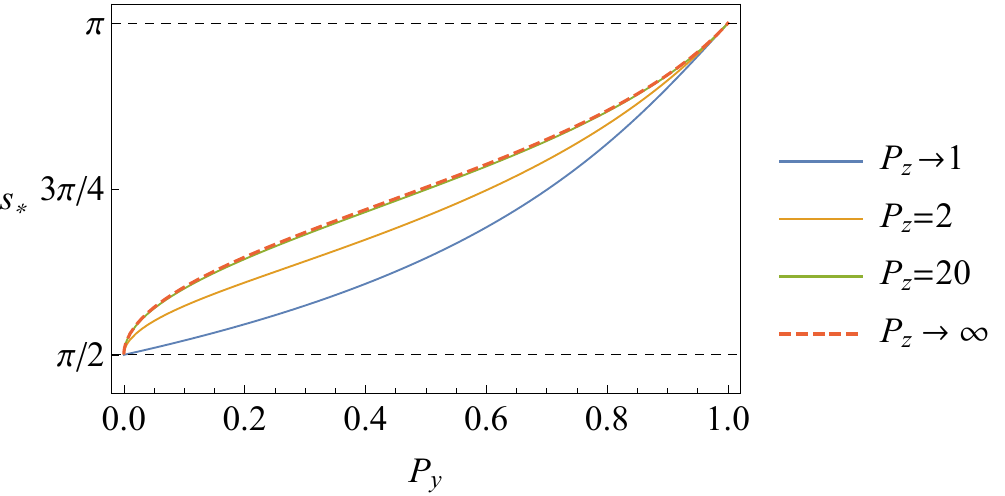}
    \caption{The location of the first conjugate point on Berger spheres (with $P_x=1$).}\label{fig:DetJ02}
\end{figure}

In general, to mimic the straight-line curve generated by a constant control Hamiltonian, we study the simple curve parametrized by  
\begin{equation}
\theta_x(s) = s \,, \quad \theta_y(s) = \theta_z(s) = 0 \,,  
\end{equation}
which corresponds to a straight line in the $\sigma_x$ direction. The geodesic deviation equation, also known as Jacobi equation, 
\begin{equation}
\frac{D^2 V^\mu}{ds^2} - R^\mu{}_{\alpha\beta\sigma} T^\alpha T^\beta V^\sigma = 0,
\end{equation}
governs the evolution of perturbations $V^\mu(s) = (V^x, V^y, V^z)$ along this geodesic. Imposing the initial condition $V^\mu(0) = 0$, the Jacobi equation can be solved analytically to yield
\begin{equation}
\begin{split}
V^x(s)  &=C_1 s
\,,\\
V^y(s)  &=  C_2 P_y(\sin (2s)-\sqrt{\frac{P_z(P_y-1)}{P_y(P_z-1)}}\sin (2As))+C_3P_z(\cos (2As)-\cos (2s))\,,\\
V^z(s)  &= C_2 P_y(\cos (2s)-\cos (2As))+C_3P_z(\sin (2s)-\sqrt{\frac{P_y(P_z-1)}{P_z(P_y-1)}}\sin (2As))\,,
\end{split}
\end{equation}
where $A = \sqrt{(P_y - 1)(P_z - 1)/(P_y P_z)}$. Conjugate points arise when a geodesic ceases to be locally minimizing. Mathematically, they correspond to the appearance of nontrivial Jacobi fields, \ie solutions to the geodesic deviation equation that vanish at both endpoints. As a result, the existence of a conjugate point corresponds to degeneracies of the Jacobi field, which is determined by the following condition
\begin{equation}
\mathcal{J}
\begin{bmatrix}
C_1 \\ C_2 \\ C_3
\end{bmatrix}
=
\begin{bmatrix}
0 \\ 0 \\ 0
\end{bmatrix}
, \quad \mathcal{J}=\begin{bmatrix}
s & 0 & 0 \\
0 & P_y\left(\sin (2s)-\sqrt{\frac{P_z(P_y-1)}{P_y(P_z-1)}}\sin (2As)\right) & P_z(\cos (2As)-\cos(2s)) \\
0 & P_y(\cos (2s)-\cos(2As)) & P_z\left(\sin(2s)-\sqrt{\frac{P_y(P_z-1)}{P_z(P_y-1)}}\sin (2As)\right)
\end{bmatrix}\,.
\end{equation}
For general $P_y$ and $P_z$, the determinant of $\mathcal{J}$ is given by
\begin{equation}\label{eq:detJ}
\det \mathcal{J}(s) = \frac{s}{2} P_y P_z \left\{ (2 + B)\cos[2(A - 1)s] + (2 - B)\cos[2(A + 1)s] - 4 \right\}\,,
\end{equation}
where $B = (2P_y P_z - P_y - P_z)/\sqrt{P_y P_z (P_y - 1)(P_z - 1)}$. 
The conjugate points are thus located at the zeros of the determinant of the Jacobi matrix, \ie  
\begin{equation}
\text{Conjugate point:} \qquad \theta_x= \theta_x(s_{\rm{conj}}) = s_{\rm{conj}} \,, \qquad \text{with} \qquad \det \mathcal{J}(s_{\rm{conj}})=0 \,. 
\end{equation}
Fig.~\ref{fig:DetJ} and Fig.~\ref{fig:DetJ02} show the locations of conjugate points for different penalty factors. 

We first reinterpret the numerical results in the main text using the single-qubit model with a homogeneous penalty factor in the odd sector,
\begin{equation}
P_x= P_y =1\,.
\end{equation}
Substituting $P_y = 1$ into eq.~\eqref{eq:detJ} reduces the determinant $\det \mathcal{J}$ to 
\begin{equation}
\det \mathcal{J}(s) \big|_{P_x=P_y}\propto \sin s \left[ P_z \sin s - (P_z - 1)s \cos s \right]\,. 
\end{equation}
It is obvious that this expression always vanishes at $s = \pi$, corresponding to conjugate points located at antipodal points (north and south poles) on the sphere. Intuitively, decreasing the penalty factor in the even sector (\ie $P_z < 1$) introduces additional conjugate points, as illustrated in Fig.~\ref{fig:DetJ}. As expected, decreasing the relative ratio $P_z$ of even and odd penalty factors moves the conjugate points closer to the initial point. Nevertheless, in the limiting case $P_z \to 0$
, the nearest conjugate point remains located at
\begin{equation}
\theta_x(s_\ast) \big|_{P_z \to 0}= \frac{\pi}{2}\,. 
\end{equation}
This behavior admits a clear geometric interpretation. In the limit $P_z \to 0$, the effective metric reduces to
\begin{equation}
\left.d s^2\right|_{P_x=P_y=1, P_z \rightarrow 0}=\cos ^2\left(2 \theta_y\right) d \theta_x^2+d \theta_y^2+\mathcal{O}\left(P_z\right) \,. 
\end{equation}
This is nothing but a two-sphere parameterized by angular coordinates $(2\theta_x\,,2\theta_y)$. As a result, the antipodal points are located at $\theta_x=\pi/2$ and $\theta_x=\pi$.  

We now return to the more general case with inhomogeneous penalties in the odd sector, \ie $P_y \neq P_x$. As discussed in the main text, a key difference from the homogeneous setup is that the penalty in the even sector should be taken to infinity for a straight-line trajectory generated by a constant control Hamiltonian to be a geodesic. In the single-qubit model, this limit corresponds to $P_z \gg P_x, P_y $. Taking the limit $P_z/P_{x,y} \to \infty$ yields a two-frequency transcendental equation
\begin{equation}
\cos (2 a s_{\rm conj}) \cos (2 s_{\rm conj})+ b \sin (2 a s_{\rm conj}) \sin (2 s_{\rm conj})=1 \,, \quad \text{with} \quad a= \sqrt{\frac{P_y-1}{P_y}} \,, b = \frac{2P_y-1}{2\sqrt{P_y(P_y-1)}} \,,
\end{equation}
which generally lacks a simple, closed-form algebraic solution. Instead, Fig.~\ref{fig:DetJ02} shows the numerical solution for the location of the first conjugate point. We find that the nearest conjugate point remains at $\theta_x(s_\ast) \ge \pi/2$, with saturation only in the limit $P_y \to 0$. To gain analytic control over this solution, we focus on the regime $P_y \ll P_x=1$. Expanding the analytic expression for $\det\mathcal{J}$ in this regime yields
\begin{equation}
 \lim_{P_z \to \infty} \det \mathcal{J}(s) \propto -2\sqrt{P_y} \sin (2 s) \sinh \left(\frac{2 s}{\sqrt{P_y}}\right)  + 4 P_y \cos (2 s) \cosh \left(\frac{2 s}{\sqrt{P_y}}\right) + \mathcal{O}(P_y)\,. 
\end{equation}
A straightforward asymptotic analysis then shows that the first conjugate point occurs at
\begin{equation}
s_\ast \approx \frac{\pi}{2} + \sqrt{P_y} \,,
\end{equation}
showing that the nearest conjugate point is located at $s_\ast \approx \frac{\pi}{2}$ in the limit $P_y \to 0$. 

In summary, the single-qubit model demonstrates that the first conjugate point associated with the straight-line geodesic generated by a constant control Hamiltonian (both in the homogeneous case and in the inhomogeneous setup with $G_{\rm even} \gg 1$) is always bounded from below by
\begin{equation}
\theta_x(s_\ast) \ge \frac{\pi}{2}\,.
\end{equation}
We have shown that this bound also holds for the SYK model under the homogeneous configuration discussed in the main text. Although a rigorous proof for general finite-dimensional systems is still lacking, it is natural to expect that this lower bound persists in more complex models such as SYK, where the operator dynamics exhibit similar geometric features.

\section{C. Small-precursor expansion of Nielsen complexity}\label{sec:appE}
Since Eq.~\eqref{eq:premainresult} is a key step in relating Krylov and Nielsen complexity, we provide here a self-contained derivation of the small-$z$ expansion of the Nielsen complexity for the precursor unitaries considered in the main text. The calculation is a local expansion around the identity element of the unitary group and is therefore closely related to the early-time expansion discussed in Ref.~\cite{Haque:2024ldr}. In the present case, however, the expansion parameter is not the physical time, but rather the precursor parameter $z$. We also compute the second order correction, showing that, in general, Krylov complexity can only provide an upper bound on Nielsen complexity.

We consider a set of Hermitian generators $\{T_I\}$ obeying
\begin{equation}
[T_I,T_J]= i f_{IJ}{}^{K} T_K ,
\end{equation}
and equip the corresponding Nielsen geometry with metric $G_{IJ}$. A circuit $U(s)$ connecting the identity to a target unitary $U_{\mathrm T}$ satisfies
\begin{equation}\label{eq:EA_app_rewrite}
\frac{dU(s)}{ds}
=
-i\,Y^I(s)T_I U(s),
\qquad
U(0)=\mathbbm{1},
\qquad
U(1)=U_{\mathrm T}.
\end{equation}
The minimizing trajectory is governed by the Euler--Arnold equation
\begin{equation}\label{eq:EA_control_rewrite}
G_{IJ}\dot{Y}^J(s)
=
G_{JL}f_{IK}{}^{J}Y^K(s)Y^L(s).
\end{equation}
For the precursor target,
\begin{equation}\label{eq:precursor_target_rewrite}
U_{\mathrm T}
=
e^{-iz\mathcal{O}(t)}
=
\exp\!\left[-iz\,\omega^I(t)T_I\right],
\end{equation}
the coefficients $\omega^I(t)$ are nonzero only along the Krylov directions, where $\omega^n(t)=\varphi_n(t)$. Since the target approaches the identity when $z\to 0$, we may solve Eqs.~\eqref{eq:EA_app_rewrite} and \eqref{eq:EA_control_rewrite} perturbatively:
\begin{equation}\label{eq:small_z_expansion_rewrite}
Y^I(s)
=
Y_0^I(s)+zY_1^I(s)+z^2Y_2^I(s)+\cdots ,
\qquad
U(s)
=
U_0(s)+zU_1(s)+z^2U_2(s)+\cdots .
\end{equation}
At zeroth order the target is simply the identity. The corresponding solution is
\begin{equation}
Y_0^I(s)=0,
\qquad
U_0(s)=\mathbbm{1}.
\end{equation}
At first order, the Euler--Arnold equation implies a constant control,
\begin{equation}
\dot{Y}_1^I(s)=0,
\qquad
Y_1^I(s)=\lambda^I .
\end{equation}
The circuit equation then gives
\begin{equation}
U_1(s)
=
-i s\,\lambda^I T_I .
\end{equation}
Matching the endpoint with
\begin{equation}
U_{\mathrm T}
=
\mathbbm{1}
-iz\,\omega^I(t)T_I
+O(z^2)
\end{equation}
fixes $\lambda^I=\omega^I(t)$. Therefore the leading Nielsen complexity is
\begin{equation}\label{eq:CN_leading_rewrite}
C_{\mathrm N}(U_{\mathrm T})
=
z\sqrt{G_{IJ}\omega^I(t)\omega^J(t)}
+O(z^3).
\end{equation}
Since $\omega^I(t)$ has support only on the Krylov directions, this becomes
\begin{equation}
C_{\mathrm N}(U_{\mathrm T})
=
z\sqrt{\sum_{m,n}G_{mn}\varphi_m(t)\varphi_n(t)}
+O(z^3).
\end{equation}
This is the leading-order expansion used in the main text. We now record the next order in the control functions, which is needed to identify the leading correction to the length. Substituting the first-order solution into the Euler--Arnold equation gives
\begin{equation}\label{eq:Y2_equation_rewrite}
G_{IJ}\dot{Y}_2^J(s)
=
f_{IK}{}^{J}\omega^K(t)G_{JL}\omega^L(t).
\end{equation}
Equivalently $\dot{Y}_2^I(s)=\Omega_2^I(t)$, where
\begin{equation}\label{eq:Omega2_rewrite}
\Omega_2^I(t)
=
G^{LI}f_{LK}{}^{J}\omega^K(t)G_{JM}\omega^M(t).
\end{equation}
The endpoint condition fixes the integration constant, giving
\begin{equation}\label{eq:Y2_solution_rewrite}
Y_2^I(s)
=
\left(s-\frac{1}{2}\right)\Omega_2^I(t).
\end{equation}
The corresponding second-order contribution to the circuit is
\begin{equation}
U_2(s)
=
-\frac{i}{2}s(s-1)\Omega_2^I(t)T_I
-\frac{s^2}{2}
\left(\omega^I(t)T_I\right)^2 .
\end{equation}
This expression determines the circuit up to second order in the precursor parameter $z$. However, it is not by itself sufficient to extract the first nontrivial correction to the Nielsen complexity. Expanding
\begin{equation}
Y^I(s)
=
zY_1^I(s)+z^2Y_2^I(s)+z^3Y_3^I(s)+\cdots ,
\end{equation}
we obtain
\begin{equation}
G_{IJ}Y^I(s)Y^J(s)
=
z^2G_{IJ}Y_1^IY_1^J
+
2z^3G_{IJ}Y_1^IY_2^J
+
z^4
\left(
G_{IJ}Y_2^IY_2^J
+
2G_{IJ}Y_1^IY_3^J
\right)
+\cdots .
\end{equation}
The $O(z^2)$ term reproduces the leading Nielsen complexity in Eq.~\eqref{eq:CN_leading_rewrite}. By contrast, the $O(z^3)$ contribution vanishes:
\begin{align}
G_{IJ}Y_1^IY_2^J
&=
\left(s-\frac{1}{2}\right)G_{IJ}\omega^I(t)\Omega_2^J(t)
\nonumber\\
&=
\left(s-\frac{1}{2}\right)
G_{IJ}\omega^I(t)
G^{LJ}f_{LK}{}^N\omega^K(t)G_{NM}\omega^M(t)
\nonumber\\
&=
\left(s-\frac{1}{2}\right)
f_{IK}{}^N\omega^I(t)\omega^K(t)G_{NM}\omega^M(t)
=
0 ,
\end{align}
since $f_{IK}{}^{N}$ is antisymmetric under $I\leftrightarrow K$, whereas $\omega^I(t)\omega^K(t)$ is symmetric. Therefore, the first nonvanishing correction to the squared length appears at order $z^4$. This correction receives contributions both from $Y_2^I$ and from the third-order control $Y_3^I$. Thus, to obtain the first nontrivial correction to the Nielsen complexity, one must also include the third-order contribution to the control functions. Combining the second- and third-order terms, the length of the minimizing path takes the form \cite{Haque:2024ldr}
\begin{equation}\label{eq:CN_z3_rewrite}
C_{\mathrm N}(U_{\mathrm T})
=
z\sqrt{G_{IJ}\omega^I(t)\omega^J(t)}
\left[
1
-
\frac{z^2}{24}
\frac{
G_{IJ}\Omega_2^I(t)\Omega_2^J(t)
}{
G_{IJ}\omega^I(t)\omega^J(t)
}
\right]
+O(z^5).
\end{equation}
Using the fact that $\omega^I(t)$ is supported on the Krylov basis, this can also be written as
\begin{equation}\label{eq:CN_z3_krylov_rewrite}
C_{\mathrm N}(U_{\mathrm T})
=
z\sqrt{\sum_{m,n}G_{mn}\varphi_m(t)\varphi_n(t)}
\left[
1
-
\frac{z^2}{24}
\frac{
G_{IJ}\Omega_2^I(t)\Omega_2^J(t)
}{
\sum_{m,n}G_{mn}\varphi_m(t)\varphi_n(t)
}
\right]
+O(z^5).
\end{equation}
The correction is negative. This is consistent with the geometric interpretation: the straight-line trajectory provides an upper bound on the Nielsen complexity, while curvature effects can only shorten the minimizing geodesic relative to this trajectory.

\end{document}